\documentclass[twocolumn,amsmath,nofootinbib]{revtex4} 
\usepackage{graphicx}
\usepackage{url}
\begin{document}
\input{epsf.sty}

\def\affilmrk#1{$^{#1}$}
\def\affilmk#1#2{$^{#1}$#2;}

\def\affilmrk#1{$^{#1}$}
\def\affilmk#1#2{$^{#1}$#2;}
\def\be{\begin{equation}}
\def\ee{\end{equation}}
\def\bea{\begin{eqnarray}}
\def\eea{\end{eqnarray}}

\title{Probing Dynamics of Dark Energy with Supernova, Galaxy Clustering and
the Three-Year Wilkinson Microwave Anisotropy Probe (WMAP) Observations
 }
\author{ Gong-Bo Zhao$^{1}$, Jun-Qing Xia$^1$,Bo Feng$^{2}$,
and Xinmin Zhang$^{1}$ }

\affiliation{${}^1$Institute of High Energy Physics, Chinese Academy
of Science, P.O. Box 918-4, Beijing 100049, P. R. China}

\affiliation{ ${}^2$ Research Center for the Early Universe(RESCEU),
Graduate School of Science, The University of Tokyo, Tokyo 113-0033,
Japan}
\date{\today.}

\begin{abstract}

Using the Markov chain Monte Carlo (MCMC) method we perform a
global analysis constraining the dynamics of dark energy in light
of the supernova (Riess "Gold" samples), galaxy clustering (SDSS
3D power spectra and SDSS lyman-$\alpha$ forest information) and
the latest three-year Wilkinson Microwave Anisotropy Probe (WMAP)
observations. We have allowed the dark energy equation of state to
get across $-1$ and pay particular attention to the effects when
incorrectly neglecting dark energy perturbations. We find the
parameter space of dynamical dark energy is now well constrained
and neglecting dark energy perturbations will make the parameter
space significantly smaller. Dynamical dark energy model where the
equation of state crosses $-1$ is mildly favored and the standard
$\Lambda$CDM model is still a good fit to the current data.

\end{abstract}


\maketitle


\setcounter{footnote}{0}

\emph{Introduction.} The recent released three year Wilkinson
Microwave Anisotropy Probe observations
(WMAP-3)\cite{Spergel:2006hy,wmap3:2006} have made so far the most
precise probe of the CMB observations. The temperature-temperature
correlation power is now cosmic variance limited up to $l \sim 400$
where the glitches on the first peak have now disappeared and the
third peak is now detected, which gives rise to a better
determination on the matter component of the
universe\cite{wmap3:2006}. In particular, the direct detection of
the CMB EE polarization spectrum and better measurements of TE
spectrum have helped a lot in the determination on the reionization
depth, which is now lower than the first year WMAP predictions and
with much smaller error bars. This has in turn helped to break the
degeneracy between the slope of the primordial scalar spectrum and
the reionization depth\cite{Spergel:2006hy,wmap3:2006}. Intriguingly
now for the fittings to the power law $\Lambda$CDM model a
Harrison-Zel'dovich spectrum is now excluded to $\sim$ $3\sigma$ by
WMAP alone, which will have profound implications in inflation if
further confirmed with higher significance
level\cite{Spergel:2006hy}. In the fittings to a constant equation
of state of dark energy, combinations of WMAP with other
cosmological constant are consistent with a cosmological constant
except for the $WMAP+SDSS$ combination, where $w<-1$ is favored a
bit more than 1$\sigma$\cite{Spergel:2006hy}. The measurements of
the SDSS power spectrum\cite{sloan} is in some sense currently the
most precise probe of the linear galaxy matter power spectrum and
will hopefully get significantly improved within the coming few
years. If the preference of $w>-1$ holds with the accumulation on
dark energy this will also help significantly on our understandings
towards dark energy. A cosmological constant, which is theoretically
problematic at present\cite{SW89,ZWS99}, will not be the source
driving the current accelerated expansion and a favored candidate
would be something like quintessence\cite{quint,pquint}. On the
other hand, the observations from the Type Ia Supernova (SNIa) in
some sense make the only direct detection of dark
energy\cite{Riess98,Perl99,Tonry03,Riess04,Riess05} and currently a
combination of $WMAP+SNIa$ or $CMB+SNIa+LSS$ are well consistent
with the cosmological constant and the preference of a
quintessence-like equation state has
disappeared\cite{Spergel:2006hy}. Intriguingly, we are also aware
that the predictions for the luminosity distance-redshift
relationship from the $\Lambda$CDM model by WMAP only are in notable
discrepancies with the "Gold" samples reported by Riess $et$
$al$\cite{Riess04}. Although the prediction by WMAP is consistent
with the measurements form the Supernova Legacy Survey
(SNLS)\cite{snls}, we are aware that the 71 high redshift type Ia
supernova alone are too weak for a cosmological probe and even when
combined with the 44 nearby SNIa their constraints on dark energy
are not yet comparable with the Riess "Gold"
sample\cite{snls,Xia:2005ge}. Although the discrepancy might be some
systematical uncertainties in the Riess "Gold" sample, this needs to
be confronted with the accumulation of the 5-year SNLS observations
and the ongoing SNIa projects like the Supernova Cosmology Project
(SCP) and from the Supernova Search Team (SST). Alternatively, this
might be due to the implications of dynamical dark energy.

The rolling scalar field of quintessence typically predicts some
evolutions of dark energy equation of state and with the
accumulation of the various kinds of observations we are now also
able to constrain the dynamics of dark energy over a constant
equation of state. Also the WMAP team has considered the constraints
on the parameter space where the equation of state is less than
$-1$. Such an equation of state is originally motivated by the
scalar field model of phantom\cite{phantom}, where there is a
negative kinetic term and typically $w$ is not a constant either for
the case of phantom and it is natural that we can expect a nonzero
$d w/ d z$ with phantom. The model of phantom has been proposed in
history due to the mild preference for a constant $w<-1$ by the
observations\cite{phantom}. Although phantom violates the weak
energy condition(WEC) and faces the dilemma of quantum
instabilities\cite{Phtproblms}, it remains possible in the
description of the nature of dark energy\cite{SPhtproblms}.

Previously with the accumulation of SNIa data and especially after
the SNIa observations from the HST/GOODS program\cite{Riess04}, many
groups have started to probe the time dependence of dark energy
equation of state \cite{sahni,Nesseris:2004wj,cooray,quintom,DES}.
Intriguingly an equation of state which crosses the cosmological
constant boundary are somewhat
favored\cite{sahni,Nesseris:2004wj,cooray,quintom}. Such a kind of
behavior is nontrivial physically since conventional quintessence
and phantom cannot realize such a behavior. In Ref.\cite{quintom} we
dubbed the new kind of dark energy $quintom$ in the sense that it
resembles the combined behavior of quintessence and phantom.
Although the model of k-essence which has non-canonical kinetic
term\cite{Chiba:1999ka,kessence} can both have a quintessence and
phantom-like behavior, as shown in
Refs.\cite{Vikman,Zhao:2005vj,Abramo:2005be} a crossing behavior is
not viable. Mathematically, the crossing reads that there exists at
least one pivot redshift, namely $z^{\star}_i$ satisfies:
\begin{equation}\label{crossing}
    \frac{dw}{dz}\mid(z^{\star}_i)\neq0,w(z^{\star}_i)=-1
\end{equation}
where $z$ denotes the redshift and $w(z)$ is the functional form of
EOS evolution. Interestingly, the quintom models differ from the
quintessence or phantom in evolution and the determination of the
fate of universe\cite{Feng:2004ff}. There exist lots of interests in
the literature presently in building of quintom-like models. For
example, with minimally coupled to gravity a simple realization of
quintom scenario is a model with the double fields of quintessence
and
phantom\cite{quintom,guozk,hu,zhang,michael,xfzhang,Wei:2005nw,cross,li}.
In such cases quintom would typically encounter the problem of
quantum instability inherited from the phantom component. However in
the case of the single scalar field model of quintom, Ref.\cite{li}
considered a Lagrangian with a high derivative term adding to the
kinetic energy and its energy-momentum tensor is equivalent to the
two-field quintom model. Such a model is theoretically possible to
resolve the problem of quantum instabilities and needs further
investigations\cite{hawking}.

Given the lack of theoretical understandings, the cosmological
observations play a crucial role to study dark energy. Probing the
dynamics of dark energy is of great significance to shed light on
theory. SNIa observations which measure the luminosity distance
depending on Hubble parameter $H(z)$, are relatively sensitive to
the dynamics of dark energy. Moreover, dark energy also leaves
imprints on Cosmic Microwave Background(CMB) through the distance to
the last scattering surface especially when we take the perturbation
of dark energy into consideration. The fluctuations of dark energy
can lessen the ISW effect and lowers the power spectrum at small
multipoles. Besides affecting $H(a)$, dark energy also modifies the
growth rate of structures via perturbation equations. Thus the LSS
data such as SDSS 3D power spectrum and Lyman-$\alpha$ forest, can
further be used in the determination of the dark energy parameter as
well as breaking the degeneracy among the various cosmological
parameters.

Given the fact that due to the problems on dark energy
perturbations\cite{Zhao:2005vj} in the conventional parametrizations
of dark energy equation of state one cannot make global fittings
with $w$ getting across the cosmological constant boundary.
Previously for example Ref. \cite{tao05} studied the perturbations
of dynamical dark energy only for the regime where $w>-1$ and Ref.
\cite{ex} considered both the cases for $w>-1$ and $w<-1$, but did
not include the perturbations for quintom like dark energy; while
other global analysis like Ref. \cite{seljak04,ex1} did not consider
dark energy perturbations. In Ref.\cite{Zhao:2005vj} we gave a
method which for the first time allows to study quintom-like dark
energy perturbations with parametrization of $w$, which resembles
the behavior of the simplest double-field quintom\footnote{Although
the multi-field dark energy models are more challenging on
theoretical aspects of naturalness, given that we know very little
on the nature of dark energy, the energy momentum of such models can
be identified with single field scalar dark energy with high
derivative kinetic terms\cite{li}. Our phenomenological formula of
perturbations on DE corresponds to such models of multi-field
(quintom) with a negligible difference around the crossing point of
-1\cite{Zhao:2005vj}. }. Our method has allowed a global analysis on
the dynamical dark energy equation of state with the observations
and like the case for a constant equation of
state\cite{WL03,Bean:2003fb}, which has also been recently shown by
Spergel $et$ $al$\cite{Spergel:2006hy}, we find incorrectly
neglecting dark energy perturbations typically make the parameter
space smaller and hence with notable bias.

 The aim of
current paper is to study the current up-to-date observational
constraints on dynamical dark energy. We extend our previous work of
Ref.\cite{Zhao:2005vj} and study the full observational constraints
on dynamical dark energy. In particular we pay great attention to
the effects of dark energy perturbations when EOS crosses -1 as in
(\ref{crossing}). Perturbations of the quintom-like models have been
studied extensively in Ref.\cite{Zhao:2005vj}. Our paper is
structured as follows: in Section II we describe the method and the
data; in Section III we present our results on the determination of
cosmological parameters with (WMAP-3)
\cite{Spergel:2006hy,wmap3:2006}, SNIa \cite{Riess04,snls} , Sloan
Digital Sky Survey 3D power spectrum (SDSS-gal) \cite{sloan} and
SDSS Lyman-$\alpha$ forrest data (SDSS-lya)\cite{lya} by global
fittings using the Markov chain Monte Carlo (MCMC) techniques~
\cite{MCMC97,MacKayBook,Neil93}; conclusions and discussions in are
presented in the last section.

\emph{Method and data}. In this section we firstly present the
general formulae of the dark energy perturbations in the full
parameter space of $w(z)$ especially for the crossing
models\ref{crossing}. In the global fitting to CMB,SN Ia and LSS
data, we adopt the parametrization of EOS as
follows:\cite{Linder:2002et}
\begin{equation}\label{linder}
    w(z)=w_{0}+w_{1}\frac{z}{1+z}
\end{equation}

\begin{figure}[htbp]
\begin{center}
\includegraphics[scale=0.75]{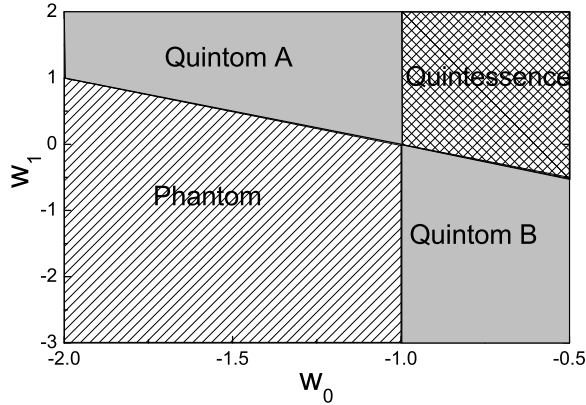}
\vskip-1.3cm \vspace{10mm}\caption{The lines $w_0=-1$ and
$w_0+w_1=-1$ divide the $w_0-w_1$ parameter space into four
parts.The gray-shaded regions are for crossing models where the
EOS of "Quintom A" models is greater than -1 in the past and
smaller than -1 today while the "Quintom B" models cross -1 in the
opposite direction.\label{fourblocks}}
\end{center}
\end{figure}
Despite our ignorance of the nature of dark energy, it is more
natural to consider the DE fluctuation whether DE is regarded as
scalar field or fluid rather than simply switching it off. The
conservation law of energy reads:
\begin{equation}\label{law}
    T^{\mu\nu}_{;\mu}=0
\end{equation}
where $T^{\mu\nu}$ is the energy-momentum tensor of dark energy
and ";" denotes the covariant differentiation. Working in the
conformal Newtonian gauge, Equation(\ref{law}) leads to the
perturbation equations of dark energy as follows \cite{ma}: \bea
    \dot\delta&=&-(1+w)(\theta-3\dot{\Phi})
    -3\mathcal{H}(\delta p/\delta\rho-w)\delta~~, \label{dotdelta}\\
\dot\theta&=&-\mathcal{H}(1-3w)\theta-\frac{\dot{w}}{1+w}\theta
    +k^{2}(\frac{\delta p/\delta\rho}{{1+w}}\delta+ \Psi)~~ ,\label{dottheta}
\eea
\bea
     \delta p&=&c^{2}_s\delta\rho + \Gamma\delta S~~. \label{soundspeed}
\eea where $c^{2}_s$ is the adiabatic sound speed and $\Gamma\delta
S$ is the entropy perturbations. For simplicity we neglect the
entropy perturbations and assume the sound speed to be
$c^{2}_s=1$\cite{Zhao:2005vj}. For the models where the EOS doesn't
cross -1, the lower left and the upper right region of
FIG.\ref{fourblocks}, the above equation(\ref{dotdelta}),
(\ref{dottheta}) is well defined. For the crossing models in
Eq.(\ref{crossing}), graphically the gray-shaded area of
FIG.\ref{fourblocks}, the perturbation equation (\ref{dotdelta}),
(\ref{dottheta}) is seemingly divergent. However basing on the
realistic two-field-quintom model as well as the single field case
with a high derivative term~\cite{Zhao:2005vj}, the perturbation of
DE is shown to be continuous when the EOS gets across -1, thus we
introduce a small positive parameter $\xi$ to divide the full range
of the allowed value of the EOS $w$ into three parts: 1) $ w> -1 +
\xi$; 2) $-1 + \xi \geq w  \geq-1 - \xi$; and 3) $w < -1 -\xi $.

For the regions 1) and 3) the perturbation is well defined by
solving Eqs.(\ref{dotdelta}), (\ref{dottheta}) as shown above. For
the case 2), the perturbation of energy density $\delta$ and
divergence of velocity, $\theta$, and the time derivatives of
$\delta$ and $\theta$ are finite and continuous for the realistic
quintom dark energy models. However for the perturbations with the
above parametrizations clearly there exists some divergence. To
eliminate the divergence typically one needs to base on the
multi-component DE models which result in the non-practical
parameter-doubling. A simple way out is to match the perturbation in
region 2) to the regions 1) and 3) at the boundary and
set\cite{Zhao:2005vj,Xia:2005ge,Xia:2006cr}
\begin{equation}\label{dotx}
  \dot{\delta}=0 ~~,~~\dot{\theta}=0 .
\end{equation}
We have numerically checked the error in the range  $|\Delta w = \xi
|<10^{-5}$ and found it less than 0.001\% to the exact multi-field
quintom model. Therefore our matching strategy is a perfect
approximation to calculate the perturbation consistently for
crossing models(\ref{crossing}). For more details of this method we
refer the readers to our previous companion papers
\cite{Zhao:2005vj,Xia:2005ge,Xia:2006cr}.

We have modified the publicly available Markov Chain Monte Carlo
package \texttt{camb/cosmomc}\cite{Lewis:2002ah}
\footnote{http://cosmologist.info/cosmomc\\http://camb.info} to
allow for the inclusion of dark energy perturbations with EOS
getting across -1\cite{Zhao:2005vj} and then sampled from the
following 8 dimensional cosmological parameter space using the
Metropolis algorithm :
\begin{equation}\label{para}
    \textbf{p}\equiv(\omega_{b},\omega_{c},\Theta_S,\tau,w_{0},w_{1},n_{s},\log[10^{10} A_{s}])
\end{equation}
where $\omega_{b}=\Omega_{b}h^{2}$ and $\omega_{c}=\Omega_{c}h^{2}$
are the physical baryon and cold dark matter densities relative to
critical density, $\Theta_S$ is the ratio (multiplied by 100) of the
sound horizon and angular diameter distance, $\tau$ is the optical
depth, $A_{s}$ is defined as the amplitude of the primordial scalar
power spectrum and $n_{s}$ measures the spectral index. We have also
marginalized over the bias factors b defined as
$b=[P_{galaxy}(k)/P(k)]^{1/2}$ which are assumed to be constant.
Basing on the Bayesian analysis, we vary the above parameters
fitting to the observational data with the MCMC method. Throughout
we assume a flat universe and
take the weak priors as: 
$\tau<0.8, 0.5<n_{s}<1.5, -3<w_{0}<3, -5<w_{1}<5 ,0.5<\Theta_S<10$,
a cosmic age tophat prior as 10 Gyr$<t_{0}<$20 Gyr. In addition, we
make use of the Hubble Space Telescope (HST) measurement of the
Hubble parameter $H_0 = 100h \quad \text{km s}^{-1} \text{Mpc}^{-1}$
\cite{freedman} by multiplying the likelihood by a Gaussian
likelihood function peaked at around $h=0.72$ with a standard
deviation $\sigma = 0.08$. We impose a Gaussian prior on the baryon
and density $\Omega_b h^2 = 0.022 \pm 0.002$ (1 $\sigma$) from Big
Bang nucleosynthesis\cite{bbn}.

In our calculations we have taken the total likelihood to be the
products of the separate likelihoods of CMB, SNIa and LSS.
Alternatively defining $\chi^2 = -2 \log {\bf \cal{L}}$, we get \be
\chi^2_{total} = \chi^2_{CMB}+ \chi^2_{SN Ia}+\chi^2_{LSS}~~~~ .\ee
In the computation of CMB we have included the three-year
temperature and polarization data with the routine for computing the
likelihood supplied by the WMAP team
\cite{Spergel:2006hy,wmap3:2006}.
 In the calculation of the likelihood from SNIa, we
use the 157 "gold" set of SNIa published by Riess $et$ $al$ in
Ref.\cite{Riess04} and marginalize over the nuisance parameter. For
LSS, we firstly use the code of \texttt{CAMB} to generate the
theoretical 3D matter power spectra of every model and fit to SDSS
3D power spectra data\cite{sloan} using the likelihood code
developed in Ref. \cite{sdssfit}. To compute the lyman-$\alpha$
forrest likelihood, we use the SDSS lyman-$\alpha$ data and
corresponding likelihood code\cite{lya} .

\emph{Results}. In this section we present our results and focus
mainly on the dark energy parameters. In particular to show the
effects of dark energy perturbations we present the resulting
constraints on the parameters for two cases simultaneously: one with
and the other (incorrectly) without dark energy perturbations.

\begin{table*}
\noindent {\footnotesize
 TABLE 1.  Mean and $1, 2\sigma$ constrains on
dark energy, spectral index and optical depth parameters using the
combined data of WMAP-3, "Gold" SNIa sample, SDSS 3D power spectra
and SDSS lyman-$\alpha$ forest information with/without DE
perturbation. Note that in the error bars of the $2\sigma$
constraints we have included the $1\sigma$ contributions of
uncertainties.

\begin{center}
\begin{tabular}{c|c|c}
  \hline
  \hline
            & \multicolumn{2}{c}{WMAP-3+RIESS + SDSS-gal + SDSS-lya} \\
   parameter& \multicolumn{1}{c|}{With Dark Energy perturbation}&\multicolumn{1}{c}{Without Dark
   Energy perturbation}\\
  \hline
  $w_0$ & $-1.146^{+0.176+0.410}_{-0.178-0.305}$ & $-1.118^{+0.152+0.324}_{-0.147-0.282}$\\
  $w_1$ & $0.600^{+0.622+0.802}_{-0.652-1.996}$ & $0.499^{+0.453+0.675}_{-0.498-1.154}$ \\
  $n_s$ & $0.962^{+0.016+0.033}_{-0.016-0.031}$ & $0.955^{+0.014+0.027}_{-0.015-0.028}$\\
  $\tau$ & $0.084^{+0.013+0.045}_{-0.012-0.046}$ & $0.079^{+0.012+0.043}_{-0.012-0.046}$ \\
  \hline
  \hline
\end{tabular}
\end{center}
}
\end{table*}

\begin{figure}[htbp]
\begin{center}
\includegraphics[scale=0.4]{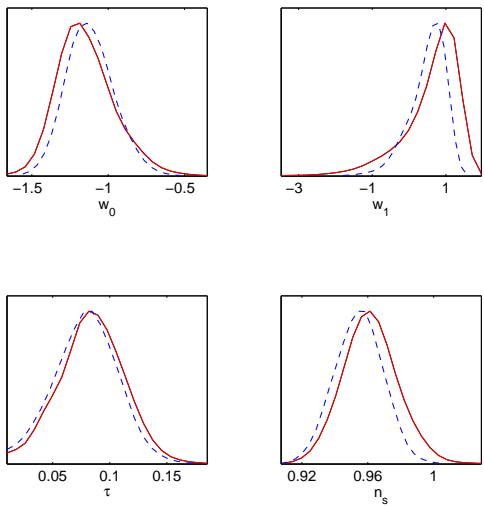}
\vskip-1.3cm \vspace{10mm}\caption{1-D constrains on individual
parameters using WMAP-3+157 "gold" SNIa+SDSS-gal+SDSS-lya with the
model of dynamical dark energy discussed in the text. Red Solid(Blue
dashed) curves illustrate the marginalized distribution of each
parameter with/without DE perturbation. \label{1d}}
\end{center}
\end{figure}

In Table 1 we list the mean and 1, 2$\sigma$ constraints on dark
energy, inflation and reionization related parameters with/without
DE perturbations. In our notations in the error bars of the
$2\sigma$ constraints we have included the $1\sigma$ contributions
of uncertainties. By virtue of the combined data and especially for
the new precision WMAP-3 data, we find that all these parameters are
well determined in our both cases. The center value of dark energy
parameters illustrate the favored dynamical DE models with EOS
crossing -1. In our framework of dynamical DE, a lower optical depth
$\tau$ is still favored which is consistent with the new results of
WMAP team\cite{Spergel:2006hy,wmap3:2006}. Moreover, a simple
scale-invariant primordial spectrum is disfavored at slightly larger
than 2-$\sigma$ even in the presence of dynamical dark energy where
the perturbations are included. In fact from Fig.\ref{1d} we find
$\tau$ is still well constrained in the present case, which breaks
the $n_S-\tau$ degeneracy and favors nontrivially a red primordial
scalar spectrum.

\begin{figure}[htbp]
\begin{center}
\includegraphics[scale=0.38]{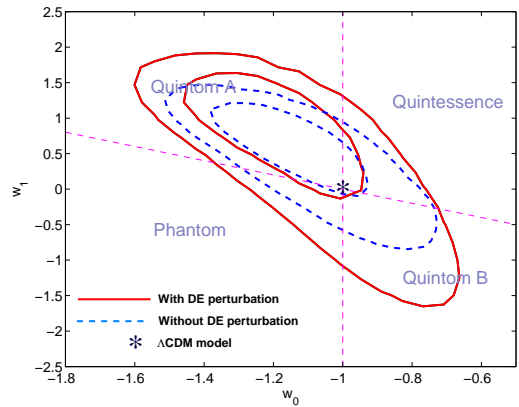}
\vskip-1.3cm \vspace{10mm}\caption{$68\%$ and $95\%$  constraints of
the 2-D contours among dark energy and the background parameters.
Red solid and blue dashed lines are for the perturbed and
(incorrectly) unperturbed DE respectively. \label{2d}}
\end{center}
\end{figure}

In Fig.\ref{1d} we delineate the corresponding posterior one
dimensional marginalized distributions of $w_0$, $w_1$, $n_s$ and
$\tau$ from our MCMC results. Since the parameters, especially for
$w_1$, are not perfectly Gaussian distributed, the mean and the
marginalized results for each parameter may be different. For the
compliment of table (1), we find the best fit values constrained by
the full dataset is $w_0=-1.363$, $w_1=1.325$, $n_s=0.964$,
$\tau=0.082$ with DE perturbation and $w_0=-1.212$, $w_1=0.839$,
$n_s=0.953$, $\tau=0.064$ with when dark energy perturbations are
(incorrectly) switched off. One can find that almost all of the best
fit parameters have been modified by the effect of DE perturbations.
Moreover, the allowed parameter space has been changed a lot and the
constraints on the background parameters have been less stringent
when including the dark energy perturbations. This can also be
clearly seen from the two dimensional contour plots in Fig.\ref{2d}.
The reason is not difficult to explain. The ISW effects of the
dynamical dark energy boosts the large scale power spectrum of
CMB\cite{Zhao:2005vj}. For a constant equation of state Ref.
\cite{WL03} has shown that when the perturbations of dark energy
have been neglected incorrectly, a suppressed ISW will be resulted
for quintessence-like dark energy and on the contrary, an enhanced
ISW is led to by phantom-like dark energy. In this sense if we
neglect dark energy contributions, there will be less degeneracy in
the determination of dark energy as well as the relative
cosmological parameters. However, dark energy perturbations are
anti-correlated with the source of matter perturbations and this
will lead to a compensation on the ISW effects, which result in a
large parameter degeneracy\cite{WL03}. In fact as we have shown that
crossing over the cosmological constant boundary would not lead to
distinctive effects\cite{Zhao:2005vj}, hence the effects of our
smooth parametrization of EOS on CMB can also be somewhat identified
with a constant effective equation of state\cite{Wang:1999fa}
\begin{equation}\label{weff}
    w_{eff}\equiv\frac{\int da \Omega(a) w(a)}{\int da \Omega(a)}~~,
\end{equation}
however the SNIa and LSS observations will break such a degeneracy
with additional geometrical constraints. Thus for the realistic
cases of including dark energy perturbations, the correlations
between the dark energy and the background parameters as well as the
auto correlations of the background cosmological parameters have
been enlarged, as can be seen from Fig.\ref{2d}.

\begin{figure}[htbp]
\begin{center}
\includegraphics[scale=0.38]{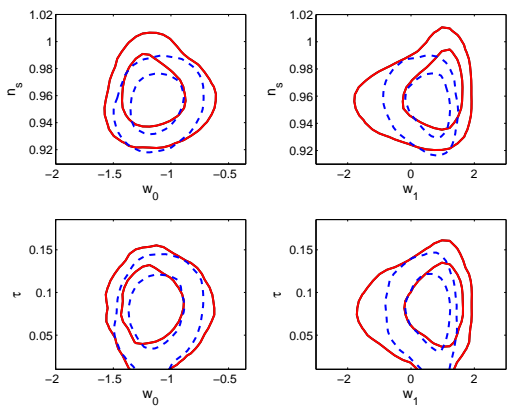}
\vskip-1.3cm \vspace{10mm}\caption{Contour plot of $w_0$ and $w_1$.
Here we use the same parametrization and data sets as FIG.2, red
solid and blue dashed lines are for perturbed and unperturbed DE
respectively. \label{corr}}
\end{center}
\end{figure}

Dark energy perturbation affect dark energy parameters most directly
and significantly, which can also be seen from Fig.\ref{corr} on the
constrains in the $(w_0, ~ w_1 )$ plane. For the parameters $( w_0 ,
~ w_1)$ the inclusion of the dark energy perturbation change its
best fit values from $(-1.212, ~~0.839)$ to $(-1.363, ~~1.325)$.
Dark energy perturbation introduces more degeneracy between $w_{0}$
and $w_{1}$ thus enlarges the contours a lot.
From the figure we can see that dynamical dark
energy with the four types are all allowed by the current
observations and Quintom A seems to cover the largest area in the
2-dimensional contours with all the data we used.

\begin{figure}[htbp]
\begin{center}
\includegraphics[scale=0.8]{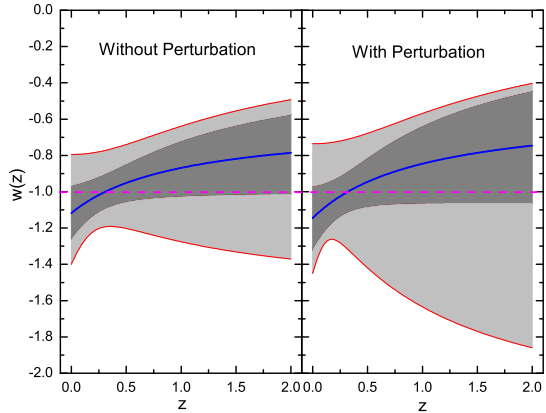}
\vskip-1.3cm \vspace{10mm}\caption{Constrains on w(z) using WMAP +
157 "gold" SNIa data + SDSS with/without DE perturbation.
Median(central line), 68\%(inner, dark grey) and 95\%(outer, light
grey) intervals of w(z) using 2 parameter expansion of the EOS in
(4).  \label{w}}
\end{center}
\end{figure}

As shown in Fig.\ref{2d} that $w_{0}$ and $w_{1}$ are in strong
correlations. The constraints on $w(z)$ are perhaps relatively model
independent, as suggested by Ref.\cite{seljak04}. Following\cite{HT}
we obtain the constraints on $w(z)$ by computing the median and 1,
2$\sigma$ intervals at all redshifts up to $z=2$. In Fig.\ref{w} we
plot the behavior of the dark energy EOS as a function of redshift
$z$, we find that at redshift $z\sim 0.3$ the constraint on the EOS
is relatively very stringent. One can see that the perturbation
reinforces the trend of DE to cross -1 at $z\sim 0.3$. However
despite in some sense $w(z)$ is well constrained by the current
data, the quintom scenario is only favored at $\sim 1 \sigma$ by the
full dataset from CMB, LSS and SNIa. The accumulation of the
observational data are still urged especially when for the probe of
the dynamical dark energy. We find the value at $z=0.3$ is
restricted at \be
 w(z=0.3) = -1.003^{+0.100+0.229}_{-0.096-0.191}
\ee
 for the case without dark
energy perturbations and \be
 w(z=0.3)= -1.008^{+0.003+0.294}_{-0.115-0.299}
\ee   when including dark energy perturbations. Correspondingly at
redshift $z=1$ the constraints turn out to be \be
 w(z=1) = -0.869^{+0.149+0.230}_{-0.160-0.407}
\ee  without perturbations and \be
 w(z=1)= -0.846^{+0.209+0.277}_{-0.221-0.789}
\ee when including dark energy perturbations. One should bear in
mind that such a constraint is not really model independent, as
shown in Refs. \cite{BCK,xia}.

\emph{Discussion and conclusion}. In this paper we have performed a
first analysis on dynamical dark energy from the latest WMAP three
year as well as the SN Ia and LSS information. Our results show that
when we include the perturbations of dark energy, the current
observations allow for a large variation in the EOS of dark energy
with respect to redshift. A dynamical dark energy with the EOS
getting across $-1$ is favored at around 1$\sigma$ with the combined
constraints from the latest WMAP , SDSS and the "gold" dataset of
SNIa by Riess $et$ $al$.

Compared with our pre-WMAP-3 results\cite{Xia:2005ge} we find now
the constraints on dark energy parameter space are improved
significantly. This lies on the fact that the CMB observations have
been improved significantly and also an inclusion of the smaller
scale power of the Lyman $\alpha$ also help a lot to break the
degeneracies. We also note in addition to the discrepancy between
WMAP-3 and the Riess sample of "Gold" SNIa, WMAP-3 also prefers a
lower value of $\sigma_8$. $\sigma_8$ is an important quantity
characterizing the size of fluctuations on the galactic scales and
this will have some profound implications on the studies of
structure formations. On the other hand Lyman $\alpha$ data prefers
a higher value of $\sigma_8$ and in some sense, our current method
of global fittings are not very strong in the probe of such
discrepancies unless all the noteworthy inconsistencies are all due
to neglecting the dynamics of dark energy. While the current WMAP-3
data and the $WMAP+SDSS$ combination favor both a deviation from
$n_S=1$ and $w=-1$, where some nontrivial dynamics might be
available "simultaneously" the both phases of accelerated expansions
more parameters characterizing the dynamics and inherent physical
quantities like the tensor to scalar ratio $r$ of the primordial
spectrum, running of the scalar spectral index $d n_S/ d \ln k$ and
also the running of $w$ as shown in the current paper need also to
be considered for both probes of the dynamics and also to avoid some
possible bias due to simply assuming no dynamics in one
sector\cite{Xia:2006cr}. Moreover in our current study we have
neglected the secondary Sunyaev-Zeldovich(SZ) effects in CMB
calculations, which can also give rise to some minor shifts on the
parameter space. But as indicated in Ref.\cite{Spergel:2006hy} such
shifts are typically small and the error bars on cosmological
parameter estimations are similar when compared with cases
neglecting the SZ effects. Further detailed analysis on the
implications of dynamical dark energy in light of the current
observations are currently in progress.

In our all results, the perturbation of dark energy plays a
significant role in the determination of cosmological parameters.
Neglecting the contributions of dark energy perturbation will lead
to biased results as shown explicitly. In the next decade, many
ongoing projects in the precise determination of cosmological
parameters will be available. We can hopefully detect the signatures
of dynamical dark energy like quintom through global fittings to the
observations, where it is crucial for us to include the
contributions of dark energy perturbations.

{\bf Acknowledgements:} We acknowledge the use of the Legacy Archive
for Microwave Background Data Analysis (LAMBDA). Support for LAMBDA
is provided by the NASA Office of Space Science. Our MCMC chains
were finished in the Shuguang 4000A system of the Shanghai
Supercomputer Center(SSC). This work is supported in part by
National Natural Science Foundation of China under Grant Nos.
90303004, 10533010 and 19925523 and by Ministry of Science and
Technology of China under Grant No. NKBRSF G19990754. We are
indebted to Patrick Mcdonald for clarifying correspondence on the
fittings to the Lyman $\alpha$ data. We thank Sarah Bridle, Antony
Lewis, Mingzhe Li, Hong Li Jun'ichi Yokoyama and PengJie Zhang for
helpful discussions and comments on the manuscript.


\vskip-0.5cm
\begin{thebibliography}{99}

\expandafter\ifx\csname
natexlab\endcsname\relax\def\natexlab#1{#1}\fi
\expandafter\ifx\csname bibnamefont\endcsname\relax
  \def\bibnamefont#1{#1}\fi
\expandafter\ifx\csname bibfnamefont\endcsname\relax
  \def\bibfnamefont#1{#1}\fi
\expandafter\ifx\csname citenamefont\endcsname\relax
  \def\citenamefont#1{#1}\fi
\expandafter\ifx\csname url\endcsname\relax
  \def\url#1{\texttt{#1}}\fi
\expandafter\ifx\csname urlprefix\endcsname\relax\def\urlprefix{URL
}\fi \providecommand{\bibinfo}[2]{#2}
\providecommand{\eprint}[2][]{\url{#2}}


\bibitem{Spergel:2006hy}
  D.~N.~Spergel {\it et al.},
  arXiv:astro-ph/0603449.


\bibitem{wmap3:2006}
L.~Page {\it et al.},
  arXiv:astro-ph/0603450.
 G.~Hinshaw {\it et al.},
  arXiv:astro-ph/0603451.
 N.~Jarosik {\it et al.},
  arXiv:astro-ph/0603452.

\bibitem{sloan}
M. Tegmark {\it et al.} (SDSS Collaboration), Astrophys.  J.   {\bf
606}, 702 (2004).



\bibitem{SW89}
S. Weinberg, Rev. Mod. Phys. {\bf 61}, 1 (1989). 

\bibitem{ZWS99}
I. Zlatev, L.-M. Wang, and P. J. Steinhardt, Phys. Rev. Lett. {\bf
82}, 896 (1999).

\bibitem{quint}
R.~D.~Peccei, J.~Sola and C.~Wetterich, Phys.\ Lett.\ B {\bf 195},
183 (1987); C. Wetterich, Nucl. Phys. B {\bf 302}, 668 (1988); C.
Wetterich, Astron. Astrophys. {\bf 301}, 321 (1995).

\bibitem{pquint}
B. Ratra and P. J. E. Peebles, Phys. Rev. D {\bf 37}, 3406 (1988);
P. J. E. Peebles and B. Ratra, Astrophys. J. {\bf 325}, L17 (1988).



\bibitem{Riess98}
A.G. Riess {\it et al.} (Supernova Search Team Collaboration),
Astron. J. {\bf 116}, 1009 (1998).

\bibitem{Perl99}
S. Perlmutter {\it et al.} (Supernova Cosmology Project
Collaboration), Astrophys. J. {\bf 517}, 565 (1999).

\bibitem{Tonry03}
J. L. Tonry {\it et al.} (Supernova Search Team Collaboration),
Astrophys. J. {\bf 594}, 1 (2003).

\bibitem{Riess04}
A.~G.~Riess {\it et al.}  (Supernova Search Team Collaboration),
Astrophys.\ J.\  {\bf 607}, 665 (2004).

\bibitem{Riess05}
A.~Clocchiatti {\it et al.}  (the High Z SN Search Collaboration),
astro-ph/0510155.


\bibitem{snls}
P. Astier {\it et al.}, astro-ph/0510447.


\bibitem{Xia:2005ge}
  J.~Q.~Xia, G.~B.~Zhao, B.~Feng, H.~Li and X.~Zhang,
  arXiv:astro-ph/0511625.



\bibitem{phantom}
R. R. Caldwell, Phys. Lett. B {\bf 545}, 23 (2002).


\bibitem{Phtproblms}
S.~M.~Carroll, M.~Hoffman and M.~Trodden, Phys. Rev. D {\bf 68},
023509 (2003); J. M. Cline, S.-Y. Jeon and G. D. Moore, Phys. Rev. D
{\bf 70}, 043543 (2004).


\bibitem{SPhtproblms}
e. g. P. H. Frampton, Phys. Lett. B {\bf 555}, 139 (2003); V. Sahni
and Y. Shtanov, J. Cosmol. Astropart. Phys. {\bf 0311}, 014 (2003);
B. McInnes, J. High Energy Phys. {\bf 0208}, 029 (2002); V.K. Onemli
and R.P. Woodard, Class. Quant. Grav. {\bf 19}, 4607 (2002); V. K.
Onemli and R. P. Woodard, Phys. Rev. D {\bf 70},107301 (2004); I. Y.
Aref'eva, A.S. Koshelev and S.Y. Vernov, astro-ph/0412619; I. Y.
Aref'eva and L. V. Joukovskaya, JHEP {\bf 0510}, 087 (2005).

\bibitem{Nesseris:2004wj}
  S.~Nesseris and L.~Perivolaropoulos,
  Phys.\ Rev.\ D {\bf 70}, 043531 (2004).


\bibitem{sahni}
U. Alam, V. Sahni and A. A. Starobinsky, J. Cosmol. Astropart. Phys.
{\bf 0406}, 008 (2004).

\bibitem{cooray}
D. Huterer and A. Cooray, Phys. Rev. D {\bf 71}, 023506 (2005).

\bibitem{quintom} B. Feng, X. Wang, and X. Zhang, Phys. Lett. B {\bf 607},
35, (2005).


\bibitem{DES}
e.g. J. Weller and A. Albrecht, Phys. Rev. Lett. {\bf 86}, 1939
(2001); Y. Wang and P. Mukherjee, Astrophys. J. {\bf 606}, 654
(2004); Y.~Wang and M.~Tegmark, Phys.\ Rev.\ Lett.\  {\bf 92} (2004)
241302; U. Alam, V. Sahni, T. D. Saini, and A. A. Starobinsky, Mon.
Not. Roy. Astron. Soc. {\bf 354}, 275 (2004); Z. H. Zhu, M. K.
Fujimoto, and X. T. He, Astron. Astrophys. {\bf 417}, 833 (2004); Y.
Gong, Class. Quant. Grav. {\bf 22}, 2121 (2005); Z.-H. Zhu,
Astrophys. J. {\bf 620}, 7 (2005); G. Chen and B. Ratra, Astrophys.
J. {\bf 612}, L1 (2004); Y. Wang, J. M. Kratochvil, A. Linde and M.
Shmakova, J. Cosmol. Astropart. Phys. {\bf 0412}, 006 (2004); W.
Godlowski and M. Szydlowski, Gen. Rel. Grav. {\bf 36}, 767 (2004);
D. Rapetti, S. W. Allen and J. Weller, Mon. Not. Roy. Astron. Soc.
{\bf 360}, 555 (2005); J. Simon, L. Verde and R. Jimenez, Phys. Rev.
D {\bf 71}, 123001 (2005); C. Csaki, N. Kaloper and J. Terning,
Annals Phys {\bf 317}, 410 (2005); E. Majerotto, D. Sapone and L.
Amendola, astro-ph/0410543; M. P. Dabrowski and T. Stachowiak,
hep-th/0411199; L. Perivolaropoulos, Phys. Rev. D {\bf 71}, 063503
(2005); Z.-H. Zhu, Astrophys. J. {\bf 620}, 7 (2005); Y. Gong and
Y.-Z. Zhang, astro-ph/0502262; S. Hannestad, Phys. Rev. D {\bf 71},
103519 (2005); G. Olivares, F. Atrio-Barandela and D. Pavon,
Phys.Rev. D {\bf 71}, 063523 (2005); W. Wang, Y.-X. Gui, S.-H.
Zhang, G.-H. Guo and Y. Shao, astro-ph/0504094; M. Szydlowski, W.
Godlowski, A. Krawiec and J. Golbiak, astro-ph/0504464; L.
Perivolaropoulos, astro-ph/0504582; X. Zhang and F.-Q. Wu,
astro-ph/0506310; B. Wang, Y. Gong and E. Abdalla, hep-th/0506069;
M.-X. Luo and Q.-P. Su, astro-ph/0506093; Z.-H. Zhu and J.S.
Alcaniz, Astrophys. J. {\bf 620}, 7 (2005); C. Csaki, N. Kaloper and
J. Terning, astro-ph/0507148; M. Ishak, A. Upadhye and D. N.
Spergel, astro-ph/0507184; D. Polarski and A. Ranquet,
astro-ph/0507290; J. S. Alcaniz and J. A. S. Lima, astro-ph/0507372;
S. Nesseris and L. Perivolaropoulos, astro-ph/0511040.


\bibitem{Chiba:1999ka}
  T.~Chiba, T.~Okabe and M.~Yamaguchi,
  Phys.\ Rev.\ D {\bf 62} (2000) 023511.

\bibitem{kessence}
C. Armendariz-Picon, V. Mukhanov and P. J. Steinhardt, Phys. Rev.
Lett. {\bf 85}, 4438 (2000); Phys. Rev. D {\bf 63}, 103510 (2001).

\bibitem{Vikman}
A. Vikman, Phys.\ Rev.\ D {\bf 71}, 023515 (2005). 

\bibitem{Zhao:2005vj}
  G.~B.~Zhao, J.~Q.~Xia, M.~Li, B.~Feng and X.~Zhang,
  Phys.\ Rev.\ D {\bf 72}, 123515 (2005).



\bibitem{Abramo:2005be}
  L.~R.~Abramo and N.~Pinto-Neto,
 astro-ph/0511562.



\bibitem{Feng:2004ff}
  B.~Feng, M.~Li, Y.~S.~Piao and X.~Zhang,
 astro-ph/0407432.

\bibitem{guozk}
Z.-K. Guo, Y.-S. Piao, X. Zhang and Y.-Z. Zhang, Lett. B {\bf 608},
177, (2005).
\bibitem{hu} W. Hu, Phys.Rev. D71 (2005) 047301. 

\bibitem{zhang} X. Zhang, hep-ph/0410292.

\bibitem{michael} R. R. Caldwell and M. Doran, Phys.\ Rev.\ D {\bf 72}, 043527
(2005).

\bibitem{xfzhang} X.-F. Zhang, H. Li, Y.-S. Piao, and X. Zhang,
astro-ph/0501652.

\bibitem{Wei:2005nw}
  For an interesting variation see H.~Wei, R.~G.~Cai and D.~F.~Zeng,
  Class.\ Quant.\ Grav.\  {\bf 22}, 3189 (2005) and for another single-field quintom model
  see e.g.  C.~G.~Huang and H.~Y.~Guo,
 astro-ph/0508171.

\bibitem{cross} I. Brevik, O. Gorbunova, gr-qc/0504001; I. Ya. Aref'eva, A.S. Koshelev, S.Yu.
Vernov, astro-ph/0507067; G.V. Vereshchagin, astro-ph/0511131;
B.M.N. Carter and I.P. Neupane, hep-th/0510109; H. Stefancic, Phys.
Rev. D71 (2005) 124036;  Z.~Chang, F.~Q.~Wu and X.~Zhang, Phys.\
Lett.\ B {\bf 633}, 14 (2006); J.~Alcaniz and H.~Stefancic,
arXiv:astro-ph/0512622; and references therein.


\bibitem{li} M. Li, B. Feng and X. Zhang, hep-ph/0503268.


\bibitem{hawking}
S. W. Hawking and  T. Hertog, Phys. Rev. D  {\bf 65}, 103515 (2002).

\bibitem{tao05}
Ch. Yeche, A. Ealet, A. Refregier, C. Tao, A. Tilquin, J.-M. Virey
and D. Yvon, astro-ph/0507170.


\bibitem{ex} P. S. Corasaniti, M. Kunz, D. Parkinson, E. J. Copeland
and  B. A. Bassett, Phys. Rev. D {\bf 70}, 083006 (2004).

\bibitem{ex1}S. Hannestad and E. Mortsell, J. Cosmol. Astropart. Phys. {\bf 0409}
001 (2004); A. Upadhye, M. Ishak and P. J. Steinhardt,  Phys.\ Rev.\
D {\bf 72}, 063501 (2005).


\bibitem{seljak04} U. Seljak {\it et al.},  Phys.\ Rev.\ D {\bf 71}, 103515
(2005). 




\bibitem{WL03}
J. Weller and  A. M. Lewis, Mon. Not. Roy. Astron. Soc. {\bf 346},
987 (2003).

\bibitem{Bean:2003fb}
For a relevant study see e.g. R.~Bean and O.~Dore,
  Phys.\ Rev.\ D {\bf 69}, 083503 (2004).






\bibitem{lya}
 P.~McDonald {\it et al.},
  arXiv:astro-ph/0407377.
\bibitem[{\citenamefont{Gamerman}(1997)}]{MCMC97}
\bibinfo{author}{\bibfnamefont{D.}~\bibnamefont{Gamerman}},
  \emph{\bibinfo{title}{Markov Chain Monte Carlo: Stochastic simulation for
  Bayesian inference}} (\bibinfo{publisher}{Chapman and Hall},
  \bibinfo{year}{1997}).

\bibitem[{\citenamefont{MacKay}(2002)}]{MacKayBook}
\bibinfo{author}{\bibfnamefont{D.~J.~C.} \bibnamefont{MacKay}}
  (\bibinfo{year}{2002}),
  \bibinfo{note}{{\url{http://www.inference.phy.cam.ac.uk/mackay/itprnn/book.html}}}.

\bibitem[{\citenamefont{Neil}(1993)}]{Neil93}
\bibinfo{author}{\bibfnamefont{R.~M.} \bibnamefont{Neil}}
  (\bibinfo{year}{1993}),
  \bibinfo{note}{{\url{ftp://ftp.cs.utoronto.ca/pub/~radford/review.ps.Z}}}.



\bibitem{Linder:2002et}
M. Chevallier and D.Polarski, Int. J. Mod. Phys. D\textbf{10},
213(2001);  E.~V.~Linder,
  Phys.\ Rev.\ Lett.\  {\bf 90}, 091301 (2003).


\bibitem{ma}
C. -P. Ma and E. Berschinger, Astrophys. J. {\bf 455} 7, (1995).

\bibitem{Xia:2006cr}
  J.~Q.~Xia, G.~B.~Zhao, B.~Feng and X.~Zhang,
  arXiv:astro-ph/0603393.




\bibitem{Lewis:2002ah}
A. Lewis and S. Bridle, Phys.  Rev.  D {\bf 66}, 103511 (2002).


\bibitem{freedman}
W. L. Freedman {\it et al.}, Astrophys.  J.  {\bf 553}, 47 (2001).

\bibitem{bbn}
S. Burles, K. M. Nollett and M. S. Turner, Astrophys.  J.  {\bf
552}, {\bf L}1 (2001).


\bibitem{sdssfit}
M. Tegmark {\it et al.} (SDSS Collaboration), Phys.  Rev.  D {\bf
69}, 103501 (2004).


\bibitem{Wang:1999fa}
e.g.  L.~M.~Wang, R.~R.~Caldwell, J.~P.~Ostriker and
P.~J.~Steinhardt,
  Astrophys.\ J.\  {\bf 530}, 17 (2000).

\bibitem{Felder:2002jk}
See e.g.  G.~N.~Felder, A.~V.~Frolov, L.~Kofman and A.~V.~Linde,
  Phys.\ Rev.\ D {\bf 66} (2002) 023507.




\bibitem{HT}
  D.~Huterer and M.~S.~Turner,
  Phys.\ Rev.\ D {\bf 64}, 123527 (2001).



\bibitem{BCK}
B. A. Bassett, P. S. Corasaniti and M. Kunz, Astrophys.  J.  {\bf
617}, {\bf L}1 (2004).


\bibitem{xia} J.-Q. Xia, B. Feng, and X. Zhang, Mod.\ Phys.\ Lett.\ A {\bf 20}, 2409
(2005).








\end{thebibliography}
\end{document}